\documentclass[preprint,showpacs,preprintnumbers,amsmath,amssymb]{revtex4}

\usepackage{txfonts}
\usepackage{amssymb}

\usepackage{graphicx}
\usepackage{dcolumn}
\usepackage{bm}
\usepackage{color}

\begin{document}


\title{Spin helical states and spin transport of the line defect in silicene lattice}

\author{Mou Yang}
\affiliation{Laboratory of Quantum Engineering and Quantum Materials, \\
School of Physics and Telecommunication Engineering,
South China Normal University, Guangzhou 510006, China}
\author{Dong-Hai Chen}
\affiliation{Laboratory of Quantum Engineering and Quantum Materials, \\
School of Physics and Telecommunication Engineering,
South China Normal University, Guangzhou 510006, China}
\author{Rui-Qiang Wang}
\affiliation{Laboratory of Quantum Engineering and Quantum Materials, \\
School of Physics and Telecommunication Engineering,
South China Normal University, Guangzhou 510006, China}

\author{Yan-Kui Bai}
\altaffiliation{Electronic address: ykbai@semi.ac.cn}
\affiliation{College of Physical Science and Information Engineering \\ 
and Hebei Advance Thin Films Laboratory, \\
Hebei Normal University, Shijiazhuang, Hebei 050024, China}

\begin{abstract}

We investigated the electronic structure of a silicene-like lattice with a line defect under the consideration of spin-orbit coupling. In the bulk energy gap, there are defect related bands corresponding to spin helical states localized beside the defect line: spin-up electrons flow forward on one side near to the line defect and move backward on the other side, and vice verse for spin-down electrons. When the system is subjected to random distribution of spin-flipping scatterers, electrons suffer much less spin-flipped scattering when they transport along the line defect than in the bulk. An electric gate above the line defect can tune the spin-flipped transmission, which makes the line defect as a spin-controllable waveguide.

\end{abstract}


\maketitle

\section{Introduction}

A variety of two-dimensional materials of the similar lattice structure as graphene have drawn intensive attention in recent years. Silicene, germanene, and stanene have the buckled honeycomb lattice, and Dirac points were found in their electronic structures.\cite{evidence} Comparing to graphene, there are advantages of these materials stemming from the lattice buckling. A normally applied electrical field induces a stagger potential and causes a band gap,\cite{gap_stagger} which is essential for the application. The buckling dramatically increases the spin-orbit coupling (SOC),\cite{gap_SOC} while it is too weak to induce observable effect in graphene.\cite{graphene_2} The SOC in a honeycomb lattice material makes it a topologic insulator and spin helical edge states exist in the edges.\cite{Kane} The combination of the stagger potential and the SOC results  in valley polarization,\cite{valley_band,anomalous} and various spin and valley related physics can be found in junction systems.\cite{junction,E_superlattice,domain} Recently, researchers successfully fabricated line defects in honeycomb lattices,\cite{control_1,control_2} which has spurred lots of discussions on their electronic properties. A tight-binding investigation reveals that the system is gapless,\cite{band} and can be regarded as a quantum waveguide.\cite{third} When a magnetic field applied, localized states arise beside the defect line, like the edge states on the edges. The quantum states around the line defect can be described by the low energy continuum model with a proper wave connection condition.\cite{wavefunction,Hall} A efficient valley filter effect can be caused by Multiple defect lines due to the valley-dependent resonance.\cite{valley_1,valley_2} The studies on the defect line are all based on graphene lattice, in which no SOC needed to be considered. Duo to the new features caused by lattice buckling, we expect more physics  can be found in the silicene-like lattice with the defect line.

\begin{figure}
\includegraphics[width=8cm]{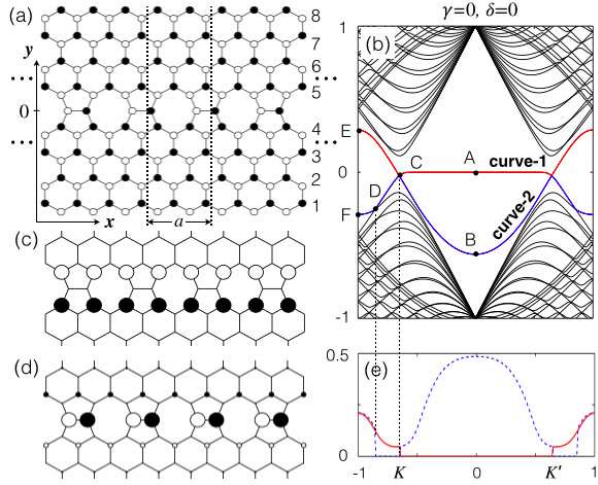}
\caption{(Color Online) \label{device} (a) $x$-$y$ projection of a silicene-like lattice with a line defect. The filled and empty circles represent atoms buckled-up and buckled-down respectively. (b) Dispersion $E(k)$ (in units of $t$) versus wavevector $k$ (in units of $\pi/a$). The lateral atom number is 40. (c) The probability distribution of the eigen state marked by point $A$ in the dispersion. Larger atom circle radius means larger probability on the atom. (d) The probability distribution of the eigen states marked by point $B$. (e) $\rho_0$ of the curve-1 (solid line) and curve-2 (dashed line) as functions of $k$.}
\end{figure}

In this paper, we investigate the band structure of a silicene-like lattice with a line defect under the consideration of spin-orbit coupling. Figure 1 (a) shows the $x$-$y$ projection of the investigated lattice, the defect atoms lie on the line $y=0$ and the buckling amplitude for the defect atoms is assumed to be the same as that of the bulk lattice. We find there are two bands related with the line defect for each spin electrons between the bulk dispersions. The states of one band are most localized on the defect atoms themselves and the other most localized on the atoms nearest to the defect line (we refer these atoms closest to the line defect as the defect edge atoms). The spin-up defect edge states propagate along one defect edge and run back along the other defect edge, and the spin-down ones behave reversely, i.e., they are spin helical states. The stagger potential makes the bands asymmetric and alters the localization properties of the defect edge states. We study the spin-flipped transport when the lattice is subjected by random distribution of spin-flipped impurities. There always exist an energy interval within the bulk gap, in which the spin-flipped transmission is much smaller than that of the bulk electrons. This property stems from the separation in real space between the states of different spins. 

\begin{figure*}\label{dispersion} 
\includegraphics[width=14cm]{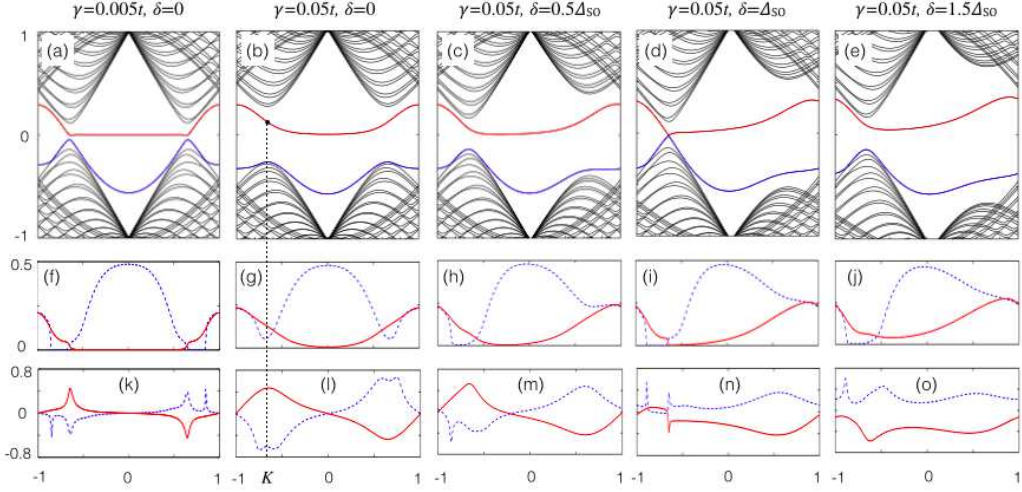}
\caption{(Color Online) 
The first row: The dispersion $E(k)$ (in units of $t$) versus $k$ (in units of $\pi/a$) for spin-up electrons. The second row: $\rho_0$ of the curve-1 (solid line) and curve-2 (dashed line) as functions of $k$. The third row: $\rho_a$ of the curve-1 (solid line) and curve-2 (dashed line) as functions of $k$. } 
\end{figure*}

\section{Calculations and Discussions}
\subsection{The Hamiltonian}
The tight-binding Hamiltonian including the SOC reads
\begin{eqnarray}\label{E}
H &=& \delta \sum_{i\alpha} \nu_{i}c_{i\alpha}^+ c_{i\alpha}  -t\sum_{\langle ij\rangle \alpha} c_{i\alpha}^+ c_{j\alpha} 
\nonumber \\
&& +i\gamma\sum_{\langle\langle ij\rangle\rangle \alpha\beta} \nu_{ij} c_{i\alpha}^+ \sigma_{\alpha\beta}^z c_{j\beta}  
\end{eqnarray} 
where $c_{i\alpha}^+$ ($c_{i\alpha}$) is the creation (annihilation) operator for an electron with spin $\alpha$ on site $i$, $\sigma^z$ is the $z$-component of Pauli matrix, and the summations with the brackets $\langle .. \rangle$ and $\langle \langle .. \rangle \rangle$ run over all the nearest and next-nearest neighbor sites, respectively. The first term is the Hamiltonian related with the stagger on-site potential, in which $\nu_i=1$ when $i$ represents a buckled up atom and $\nu_i=-1$ for buckled down atom, and $\delta$ is the stagger potential amplitude. The second term is the Hamiltonian of the nearest neighbor hopping with hopping energy $t$. The third term is the SOC Hamiltonian which involves the next-nearest neighbor hopping with amplitude $\gamma$ and a path dependent amplitude $\nu_{ij}$. For the electron couples form atom $i$, mediated by a nearest neighbor site and to a next-nearest neighbor atom $j$, we have $\nu_{ij}=1$ if it makes a left turn and $\nu_{ij}=-1$ if goes a right turn. Since the line defect lies along $x$-direction, and the wavevector in $x$-direction is a good quantum number. The calculation is conducted in a translational cell. In $y$-direction, the periodical edge condition is adopted to avoid the distraction of the edge states, which are not our targets.

\begin{figure}
\includegraphics[width=7cm]{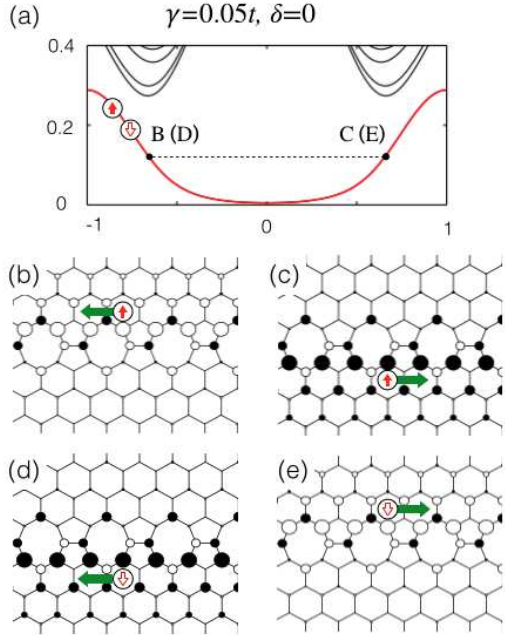}
\caption{(Color Online) 
(a) A zoom picture of the dispersion for spin-up electrons (the dispersion for the spin-down electrons is exactly the same). (b) and (c) The probability distribution of states labeled by point $B$ and $C$ in (a) for the spin-up band respectively. (d) and (e) The probability distribution of states labeled by point $D$ and $E$ for the spin-down band respectively. }
\end{figure}

\subsection{The Basic Case: $\delta=0$ and $\gamma=0$}
Firstly, we investigate the electronic structure of the system when both the stagger potential and the SOC are turned off (i.e., $\delta=0$ and $\gamma=0$). The dispersion is shown as Fig. \ref{device} (b). It can be seen that the dispersion is quite similar to that of a graphene ribbon, except that there are two additional curves lacking of electron-hole symmetry. The two bands are labeled by curve-1 and curve-2 in the figure and we will conduct detailed investigation on their properties for a variety of parameters. There is a flat part on curve-1, which implies that these states bear analogous properties of edge states. Figure. 1 (c) shows the electron probability distribution of the eigen state represented by point $A$ on curve-1, and one can see that the density is most localized on the defect edge atoms. If we choose another point on the flat band apart from point $A$ (a point between $A$ and $C$) to study, we find the density decays away from the defect edge atoms to the bulk of either side, and the decay rate depends on the deviation of the point studied from point $A$ (not shown in the figure). These features are just those of edge states for a zigzag graphene ribbon, which is not strange because the defect edge atoms are just the real edge ones if the defect atoms are removed. For this reason, we call these defect-nearest atoms as defect edge ones. The electron density of point $B$ on curve-2 is shown in Fig. 1 (d). The density is most localized on the defect atoms and slightly scattered on nearby atoms. For other points on the same curve near point $B$, the densities are more scattered on more atoms around the defect atoms, and decay away into the bulk (not shown in the figure). We also examine the properties of other parts of curve-1 and curve-2. The states of $EC$ and $FD$ are distributed on both the defect atoms and the defect edge atoms, as the states of part $CB$ of curve-2, while, $DC$ represents bulk states, which is result of the band crossing that occurs at point $D$. The density distributions of the two bands are symmetric with respect to the defect line. 

To describe the localization on the defect atoms, we define quantity
\begin{eqnarray}
\rho_0 &=\rho(y=0),
\end{eqnarray}
where $\rho$ is the probability distribution. Figure 1 (e) shows $\rho_0$ of curve-1 and curve-2 as functions of $k$. From $E$ to $C$, $\rho_0$ decreases continuously, abrupt change happens at point $C$ because of the band crossing, and it vanishes for part $CA$ (the flat part) since it represent defect edge states. $\rho_0$ of curve-2 experiences one more abrupt change because there is an additional crossing at point $D$ besides of the crossing at point $C$; part $FD$ of it is almost overlap with $\rho_0$ curve for curve-1 since they have similar localization properties; for part $DC$, $\rho_0$ is zero, which reflects the properties of the bulk states and the electron probability on the defect atoms is infinitesimal; and from $C$ to $B$, the density on defect atoms is continuously increased.

\subsection{The General Cases}
Now we turn the SOC term on to a small value. For this case the system is spin-dependent, we only study the properties for the spin-up electrons for now, and discuss spin-down electrons later. The SOC induces a gap $2\Delta_{SO}$ at each valley for a perfect bulk silicene-like lattice, which depends on the SOC by
\begin{eqnarray}
\Delta_{SO}=3\sqrt3 \gamma.
\end{eqnarray}
The small gap can be found between curve-1 and  curve-2 near valley $K$ in Fig. 2 (a), and it causes slightly smearing of the abrupt changes of $\rho_0$ versus $k$ for both curve-1 and curve-2, as illustrated in Fig. 2 (f). When we increase the SOC amplitude, the gap at valley $K$ increases correspondingly, the smearing of $\rho_0$ is more apparently, the defect states around point $C$ on curve-1 and the bulk states around the point on curve-2 are mixed with each other, and the $\rho_0=0$ part disappears. The bulk gap at valley $K$ or $K'$ is not the real gap between curve-1 and curve-2, because the bottom of curve-1 remains almost unchanged at $E=0$ when $\gamma$ changes.

In the energy gap, the SOC drives spin-up electrons piled up at one edge if edges exist, and spin-down electrons at the other edge. Because the defect line can be regarded as another type of edge, we expect electrons with different spins accumulate near different defect edges even no real edge exist (periodic edge condition in $y$-direction is adopted, so no real edge exists) and the electron probability distribution exhibits asymmetry with respect to the defect line. To describe the asymmetry, we define the quantity
\begin{eqnarray}
\rho_a &=\int_{y>0}\rho dy - \int_{y<0}\rho dy.
\end{eqnarray}
Figure 2 (k) shows $\rho_a$ of curve-1 and curve-2 as functions of $k$. A small SOC can induces a remarkable asymmetry of $\rho_a$ (i.e., $\rho_a \neq 0$) near valley $K$, and a larger SOC $\Delta=0.05t$ results in the asymmetry of the whole interval of $k$. Around valley $K$ the electron probability corresponding to curve-1 is localized near the upper defect edge ($\rho_a>0$) and near the lower defect edge ($\rho_a<0$) around the other valley, and for the curve-2 situation is reversed.

Next we switch on the stagger potential amplitude $\delta$. The combination of the stagger potential and SOC term causes two different energy gaps, $2\Delta_K$ and $2\Delta_{K'}$, for valleys $K$ and $K'$ for a perfect silicene-like lattice, respectively, which is called as valley polarization effect and only happens for $\Delta_{SO}\delta \neq 0$. The two gaps are determined by
\begin{eqnarray}
\begin{split}
\Delta_K &=|\delta - \Delta_{SO}|, \\
\Delta_{K'} &=|\delta + \Delta_{SO}|.
\end{split}
\end{eqnarray} 
For small stagger potential $\delta<\Delta_{SO}$, the gap of valley $K$ shrinks and that of valley $K'$ enlarges, which makes the bulk dispersions around valleys $K$ and $K'$ not identical, breaks the left-right symmetry of $\rho_0$, and spoils the anti-left-right symmetry of $\rho_a$, as illustrated in Fig. 2 (c), (h), and (m). When $\delta$ is increased to $\delta = \Delta_{SO}$, we have $\Delta_K=0$, saying, the gap of valley $K$ vanishes and the crossing between curve-1 and curve-2 returns. The abrupt change of $\rho_0$ at the crossing point is revivified. The asymmetry of the electron probability around valley $K$ is almost unobservable while remains apparent around valley $K'$. These features can be seen in Fig. 2 (d), (i), and (n).
When we increase $\delta$ so as to $\delta > \Delta_{SO}$, we have two nonidentical gaps for two different valleys again. Fig. 2 (e), (j), and (o) show the dispersion, $\rho_0$, and $\rho_a$ for this case.
The bulk dispersion and the $\rho_0$ curve are quite similar to those for the case of $\delta< \Delta_{SO}$, but the bottom of curve-1 is above $E=0$ and no more $\rho_0=0$ point can be found on the curve of $\rho_0$ for curve-1. In the whole interval of $k$, $\rho_a<0$ for curve-1, which is very different from the case $\delta < \Delta_{SO}$.

If we only consider the spin-down electrons, all the curves appearing in Fig. 2 are flipped left-to-right with these symmetric curves keep intact.

Fig 3 (a) shows the energy dispersions for both spin-up and spin-down for $\Delta_{SO} \neq 0$ and $\delta=0$. The curves for spin-up and spin-down overlap completely, but the probability distribution for the two defect edge bands are different. We study the states represented by four points at the same energy, say, points $B$ and $C$ on curve-1 for spin-up and $D$ and $E$ for spin-down. The probability distribution corresponding to the four point is illustrated in Fig. 3 (b) through (e). At the energy, the spin-up electrons go right along the lower defect edge and flow left along the upper defect edge, in other words, run anti-clockwise around the defect line, while the spin-up electrons go clockwise.  

\begin{figure}
\includegraphics[width=7cm]{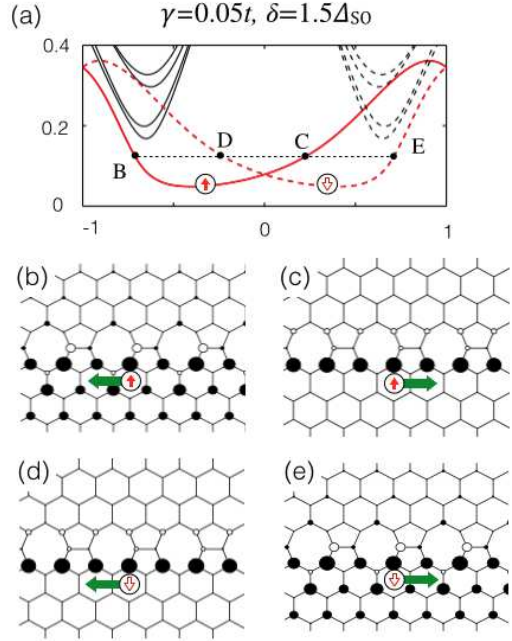}
\caption{(Color Online)  (a) A zoom picture of the dispersions for spin-up (solid lines) and spin-down (dashed lines) electrons. (b), (c), (d), and (e) The probability distribution of states labeled by points $B$, $C$, $D$, and $E$ in (a), respectively.}
\end{figure}

When both SOC and stagger potential are present, the energy dispersions for spin-up and spin-down electrons are not identical, as shown in Fig 4 (a). The probability distribution of four point $B$ through $E$ labeled in Fig. 4 (a) are illustrated in Fig. 4 (b) through (e). At the energy represented by the horizontal line, electrons with both spins move along the lower defect edge leaving the upper defect edge almost empty. This can be understood by that atoms belong to different defect edges are buckled reversely and a electric field across the defect line. Because $\delta>\Delta_{SO}$ is adopted in the figure, the electric field is dominate and drives the electron probability on one side of the defect line.

\subsection{Spin Dependent Transport}

\begin{figure}
\includegraphics[width=6.5cm]{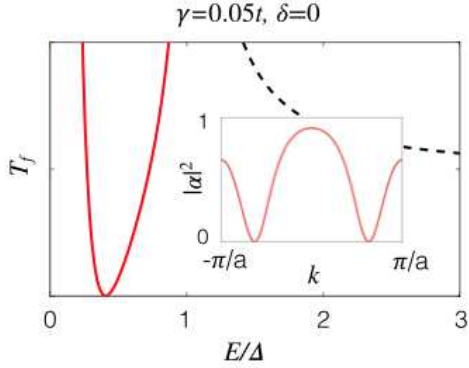}
\caption{(Color Online) \label{figT} Spin-flipped transmission $T_f$ (in arbitrary units) as a function of energy for both the defect line guided transport (solid line) and $x$-direction propagation for bulk electrons (dashed line). The inset shows the curve of $|\alpha|^2=|\langle \phi_{k\downarrow}|\phi_{k\uparrow}\rangle|^2 $ as a function of $k$.}
\end{figure}

The spacial separation of the electrons with different spins is helpful for the suppression of spin flip when electrons propagating. A small bias along the defect line drives electrons from the left going to the right. If the electron is injected at the energy within the bulk energy gap and above the bottom of curve-1, the transport is supported by the defect edge states. When no stagger potential is applied, the spin-up and spin-down electrons flow rightward alone the lower and upper defect edges, respectively (see Fig. 3). We consider there are some spin-flipping scatterers random distributed on the plane, and the scatterers are assumed to be sufficiently weak, microscopically large but macroscopically small, and identical with each other. The electrons cannot be scattered back because back scattering needs large $k$ transfer, but the spin can be flipped without change of $k$. This type of scatterers can be regarded as a simple model of the local lattice distortion induced by alien atom absorption.\cite{hydro} The local distortion, for example, a bump, leads to spin-dependent scattering and is relaxed within the area of the dimension of tens of lattice constant.\cite{spread} Besides the lattice distortion, the on-site energy of the alien atom is a short-ranged scatterer, which is spin-independent and not our aim. 
In frame of first order perturbation theory, the spin-flipped transmission (the detailed derivation is given in the Appendix) is
\begin{eqnarray} \label{Tf}
T_{f} &=& 4\pi^2\eta^2 \cdot \rho(E_{k\downarrow})  \left|\langle \phi_{k\downarrow}|\phi_{k\uparrow}\rangle \right|^2 \rho(E_{k\uparrow}) \nonumber \\
&=& \eta^2 v^{-2}|\alpha|^2,
\end{eqnarray}
where $\rho(E)$ is density of states at energy $E$, $v=\partial E_k/\partial k=(2\pi\rho)^{-1}$ is the velocity ($\hbar=1$ is adopted), and $\alpha=\langle \phi_{k\downarrow}|\phi_{k\uparrow}\rangle$ is the overlap of lateral eigen states of the spin-up and spin-down electrons, and $\eta=Vns/S$ is a parameter to characterize 2D disorder with $n$, $s$ and $V$ being the number, area, and potential of scatterer and $S$ the area of the sample. For the bulk electrons going parallel to the $x$-direction, the spin-flipped transmission is described by the same equation. The dispersion relation of bulk electrons is $E^2=\Delta^2+v_F^2k^2$, so we have the velocity square is $v^2=v_F^2\cdot(E^2-\Delta^2)/E^2$, where $v_F$ is the Fermi velocity of perfect silicene, and $\alpha=1$ for bulk electrons. Figure \ref{figT} shows the spin-flipped transmission as a function of energy for both the defect line guided transport and $x$-direction propagation for bulk electrons. Above the gap, the flipped transmission decreases with increasing energy and tends to a stabilized value because $v\rightarrow v_F$ when $E\gg\Delta$. However, within the bulk gap, there exists an interval in which the spin-flipped transmission is smaller than that for the bulk state, this is because $\alpha$ vanishes at some wavevectors near valleys $K$ and $K'$, as shown in the inset of Fig. \ref{figT}, and thus $\alpha=0$ occurs a certain energy.

The spin-flipped feature of the line defect results in interesting spin transport effects. If we tune the system that the Fermi energy is aligned at the flip-free point, the line defect works as an ideal spin guide; when we place a top gate (the gate length along the line defect must be limited within the coherence length) to push the Fermi energy away from the flip-free point, we have the out-coming electrons with the spin being a superposition of spin-up and spin-down, and the weights of different spins depends on the length and potential of the gate. A precision prediction of the weights beyond the first order perturbation, but the qualitative physics can be expected as that.

To our knowledge, silicene can now only be grown on metal surfaces,\cite{review} which is an obstacle to perform transport measurement, and long and regular defect line in silicene has not being reported experimentally yet. However, our calculation for the spin transport is also applicable for the graphene defect line if the SOC cannot be neglected, which was recently successfully to enhanced to a few meV by weak hydrogenating.\cite{hydrogenate}

\section{Summary}

We studied the electronic structure of silicene with a line defect. Spin helical states was found around the line defect. These states belong to energy bands in the bulk energy gap. When the lattice is subjected by a distribution of spin-flipping scatterers, within the bulk gap the spin-flipped transmission is much smaller that of the bulk electrons and can be controlled by a gate above the defect line.

\acknowledgements
This work was supported by NSF of China Grant No. 11274124, 
No. 11474106, No. 11174088, and Hebei NSF Grant No. A2012205062.

\appendix
\section{Derivation of Equation (\ref{Tf})}

Supposing the initial wavefunction is denoted as $\psi_0$, in frame of first perturbation theory, the scattered wavefunction can be written as
\begin{eqnarray}
\psi' &=& G_0V\psi_0
\end{eqnarray}
where $G_0$ is the Green's function of scattering free system and $V$ is the scattering potential operator. If the wavefunction is initialized as a plane wave $e^{ikx}$, the scattered wavefunction is 
\begin{eqnarray}
\psi' =\int \frac{dk'}{2\pi}e^{ik'x}\frac{1}{E_{k}-E_{k'}}V_{k'k}
\end{eqnarray}
where $E_{k/k'}$ is the energy of the wavefunction with wavevector $k/k'$ for the unperturbed system.
In the first Brillouin zone, we can find even number of poles (or solutions of $k'$) on the real axis of $k'$ by solving the equation $E(k') = E_k$, and we denote the solutions corresponding to positive velocity as $\kappa^+$ and those corresponding to negative as $\kappa^-$. The integration over $k'$ can be transform into contour integration around these poles, 
\begin{eqnarray} \label{contour}
\int \frac{dk'}{2\pi} &=& \left\{ 
\begin{split}
\sum_i\ointctrclockwise_{\kappa_i^+} \frac{dk'}{2\pi}, \quad x>0\\
\sum_i\ointclockwise_{\kappa_i^-} \frac{dk'}{2\pi}, \quad x<0
\end{split}
\right.
\end{eqnarray}
For both cases $x>0$ and $x<0$, each contour integration of $(E_{k'}-E_k)^{-1}$ leads to the result $|dk'/dE|_{k'=\kappa}=v_\kappa^{-1}$, so the scattered wavefunction is
\begin{eqnarray}
\psi'_{k'} &=& -iv_{k'}^{-1}e^{ik'|x|}V_{k'k}
\end{eqnarray}
where $k'$ means the final wavevector which satisfying $E_{k'}=E_k$, saying, the energy conservation. The coefficient of the scattered plane wave is
\begin{eqnarray}
-iv_{k'}^{-1}V_{k'k}
\end{eqnarray}
The process is initialized with the flux $\sim v_k$, and scattered into the state of the flux $\sim v_{k'}$. The reflection or transmission component (if the final velocity is coincide with the initial one, transmission, otherwise, reflection) from $k$ to $k'$ is associated with the process is  

\begin{eqnarray} \label{T}
T_{k'k} &=& v_{k'}\left|v_{k'}^{-1}V_{k'k}\right|^2v_k^{-1} 
=v_{k'}^{-1}\left|V_{k'k}\right|^2v_k^{-1} \nonumber  
\\
&=& 2\pi\rho(E_{k'}) \cdot\left|V_{k,k'}\right|^2 \cdot2\pi\rho(E_k)
\end{eqnarray}
where $\rho=dn/dE=dn/dk\cdot dk/dE=(2\pi v)^{-1}$ is the density of states at the energy $E$.

The derivation is based on one-dimension spinless system, but it can be extended straightforward to multidimensional spinful system by regarding the wavevector as a new composite quantum number, $k\rightarrow(\boldsymbol{k},\sigma)$, where $\sigma=\uparrow,\downarrow$. For one-dimensional spinless system $k'\neq k$ must be met to ensure that the final state is not identical to the initial state, but for the spinful system, this restriction is not necessary. By setting the stats $(k,\uparrow)$ and $(k,\downarrow)$ be the initial and final states, respectively, we have Eq. (\ref{Tf}).


\begin{references}

\bibitem{evidence} L. Chen, C.-C. Liu, B. Feng, X. He, P. Cheng,  Z. Ding, S. Meng, Y. Yao, and K. Wu, Phys. Rev. Lett. {\bf109}, 056804 (2012).

\bibitem{gap_stagger} N. D. Drummond, V. Z\'{o}lyomi, and V. I. Fal'ko, Phys. Rev. B {\bf85}, 075423 (2012).

\bibitem{gap_SOC} C.-C. Liu, W. Feng, and Y. Yao, Phys. Rev. Lett. {\bf107}, 076802 (2011).

\bibitem{graphene_2} A. H. C. Neto, F. Guinea, N. M. R. Peres, K. S. Novoselov, and A. K. Geim, Rev. Mod. Phys. {\bf81}, 109 (2009).

\bibitem{Kane} C. L. Kane and E. J. Mele, Phys. Rev. Lett. {\bf95}, 226801, (2005).

\bibitem{valley_band} M. Ezawa, Phys. Rev. Lett. {\bf109}, 055502 (2012).

\bibitem{anomalous} H. Pan, Z. Li, C.-C. Liu, G. Zhu, Z. Qiao, and Y. Yao, Phys. Rev. Lett. {\bf112}, 106802 (2014).

\bibitem{junction} T. Yokoyama, New J. Phys. {\bf16}, 085005 (2014).

\bibitem{E_superlattice} S. K. Wang, J. Wang, and K. S. Chan, New J. Phys. {\bf16}, 045015 (2014).

\bibitem{domain} Y. Kim, K. Choi, J. Ihm, and H. Jin, Phys. Rev. B {\bf89}, 085429 (2014).

\bibitem{control_1} J. Lahiri, Y. Lin, P, Bozkurt, I. I Oleynik, and M, Batzill, Nature nanotech. {\bf5}, 326-329 (2010).

\bibitem{control_2} J.-H. Chen, G. A\`{u}tes, N. Alem, F. Gargiulo, A. Gautam, M. Linck, C. Kisielowski, O. V. Yazyev, S. G. Louie, and A. Zettl, Phys. Rev. B {\bf89}, 121407(R) (2014).

\bibitem{band} J. Song, H. Liu, H. Jiang, Q.-f. Sun, and X. C. Xie, Phys. Rev. B {\bf86}, 085437 (2012).

\bibitem{third} D. A. Bahamon, A. L. C. Pereira, and P. A. Schulz, Phys. Rev. B {\bf83}, 155436 (2011).

\bibitem{wavefunction} L. Jiang,G. Yu, W. Gao, Z. Liu, and Y. Zheng, Phys. Rev. B {\bf86}, 165433 (2012).

\bibitem{Hall} H.-B. Yao, X.-L. Lu, and Y.-S. Zheng, Phys. Rev. B {\bf88}, 235419 (2013).

\bibitem{valley_1} D. Gunlycke and C. T. White, Phys. Rev. Lett. {\bf106}, 136806 (2011).

\bibitem{valley_2} Y. Liu, J. Song, Y. Li, Y. Liu, and Q.-f. Sun, Phys. Rev. B {\bf87}, 195445 (2013).

\bibitem{hydro} M. Gmitra, D. Kochan, and J. Fabian, Phys. Rev. Lett {\bf110} (2013).

\bibitem{spread} A. V. Savin and Y. S. Kivshar, Phys. Rev. B {\bf88}, 125417 (2013).









\bibitem{review} A. Kara, H. Enriquez, A. P. Seitsonen, L.C. Lew Yan Voon, S. Vizzini, B. Aufray, H. Oughaddou, Surf. Sci. Rep. {\bf67}, 1 (2012).
\bibitem{hydrogenate} J. Balakrishnan, G. K. Koon, M. Jaiswal, A. H. Castro Neto, and B. \"{O}zyilmaz, Nat. Phys. {\bf9}, 284 (2013).

\end{references}
\end{document}